\documentclass[aps,prl,superscriptaddress,twocolumn,showpacs]{revtex4}

\usepackage{graphicx} 
\usepackage{dcolumn}  
\usepackage{bm}       

\usepackage{amsmath}
\usepackage{xspace} 
\def\PD      {\ensuremath{D}\xspace}     
\def\Dbar    {{\kern 0.2em\overline{\kern -0.2em \PD}{}}\xspace}

\def\DorDbar {\kern 0.18em\optbar{\kern -0.18em D}{}\xspace}

\def\Dzb     {{\ensuremath{\Dbar{}^0}}\xspace}            
\def\PK      {\ensuremath{K}\xspace}    
\def\Kbar    {{\kern 0.2em\overline{\kern -0.2em \PK}{}}\xspace}

\def\KorKbar {\kern 0.18em\optbar{\kern -0.18em K}{}\xspace}

\begin{document}


\title{\boldmath Measuring $CP$ violation and mixing  in charm \\ with inclusive self-conjugate multibody decay modes}




\author{S.~Malde}
\author{C.~Thomas}
\affiliation{University of Oxford, Denys Wilkinson Building, Keble Road,  OX1 3RH, United Kingdom}
\author{G.~Wilkinson}
\affiliation{University of Oxford, Denys Wilkinson Building, Keble Road,  OX1 3RH, United Kingdom}
\affiliation{European Organisation for Nuclear Research (CERN), CH-1211, Geneva 23, Switzerland}


\date{\today}

\begin{abstract} 
\noindent
Time-dependent studies of  inclusive charm decays to  multibody self-conjugate final states can be used to determine the indirect $CP$-violating observable $A_\Gamma$ and the mixing observable $y_{CP}$, provided that the fractional $CP$-even content of the final state, $F_+$, is known. This approach can yield significantly improved sensitivity compared with the conventional method that relies on decays to $CP$ eigenstates.  In particular, $D \to \pi^+\pi^-\pi^0$ appears to be an especially powerful channel, given its relatively large branching fraction and the high value of $F_+$ that has recently been measured at charm threshold.  
%
\end{abstract}

\maketitle


It is of great interest to search for effects of indirect $CP$ violation in time-dependent studies of neutral charm-meson decays. In the Standard Model indirect $CP$ violation is expected to be well below the current level of experimental precision~\cite{THEORY1}, but many models of New Physics predict enhancements~\cite{THEORY1b}.   A very  important $CP$-violating observable is $A_\Gamma$, which is measured  from the difference in lifetimes of the decays of $D^0$ and \Dzb mesons to a $CP$ eigenstate.  In this paper it is shown how inclusive self-conjugate multibody decays that are not $CP$ eigenstates can also be harnessed for the measurement of $A_\Gamma$, provided that their fractional $CP$-even content, $F_+$,  is known.  This new approach has the potential to improve significantly the knowledge of $A_{\Gamma}$
and has become possible thanks to measurements of $F_+$  that have recently begun 
to emerge from analyses of coherent charm-meson pairs produced at the $\psi(3770)$ resonance~\cite{MINAKSHI,FOURPI}. 
Furthermore, it is explained how exploiting these decays can also provide a corresponding improvement in the precision on $y_{CP}$, which is an important observable that describes $D^0 \Dzb$ oscillations.
For the purpose of concreteness the discussion is presented for the example decay $D \to \pi^+\pi^-\pi^0$, although the results are valid for all self-conjugate multibody modes.  Here and throughout the discussion $D$ indicates a neutral charm meson; this notation is used when it is either unnecessary or not meaningful to specify a flavour eigenstate.

\vspace{0.2cm}
\noindent
{\bf \boldmath Measurements with $CP$ eigenstates}
\vspace{0.2cm}

In the $D$-meson system the mass eigenstates, $D_{1,2}$, are related to the flavour eigenstates $D^0$ and \Dzb as follows:
\begin{equation}
|D_{1,2}\rangle = p|D^0\rangle \pm q|\Dzb\rangle,
\end{equation}
where the coefficients satisfy $|p|^2 + |q|^2 = 1$ and 
\begin{equation}
r_{CP} e^{i \phi_{CP}} \equiv \frac{q}{p}.
\label{eq:cpvdefinition}
\end{equation}
The phase convention $CP|D^0 \rangle = |\Dzb \rangle$ is adopted. 
Indirect $CP$ violation occurs if $r_{CP} \ne 1$ and/or $\phi_{CP} \ne 0$.
Charm mixing is conventionally parameterised by the quantities $x$ and $y$, defined as
\begin{equation}
x \equiv \frac{M_1 - M_2}{\Gamma},  \qquad y \equiv \frac{\Gamma_1 - \Gamma_2}{2\Gamma},
\label{eq:xydefinition}
\end{equation}
where $M_{1,2}$ and $\Gamma_{1,2}$ are the mass and width of the two neutral meson mass eigenstates, and $\Gamma$ the mean decay width of the mass eigenstates.  In the chosen convention $D_1$ is almost $CP$ even. The average of currently available measurements gives $x=(0.41^{+0.14}_{-0.15})\%$ and
$y =(0.63^{+0.07}_{-0.08})\%$~\cite{HFAG}.

Consider an environment where charm mesons are produced incoherently, such as the LHC or an $e^+e^-$ $B$-factory, and are observed through their decay into a $CP$ eigenstate of eigenvalue $\eta_{CP}$.  Time-dependent measurements allow the decay widths $\hat{\Gamma}$  and $\hat{\bar{\Gamma}}$ to be determined for mesons produced in the $D^0$ and \Dzb flavour states, respectively.  From these quantities the $CP$-violating observable $A_\Gamma$ and mixing observable $y_{CP}$ may be constructed:
\begin{equation}
A_\Gamma \equiv \eta_{CP} \frac{ \hat{\Gamma} - \hat{\bar{\Gamma}} }{\hat{\Gamma} + \hat{\bar{\Gamma}} }, \qquad 
y_{CP} \equiv \eta_{CP} \left( \frac{\hat{\Gamma} + \hat{\bar{\Gamma}} }{2\Gamma}  - 1 \right) .
\label{eq:agammaycp}
\end{equation}
Assuming $x$, $y$, $(r_{CP} - 1/r_{CP})$ and $\phi_{CP}$ to be small, and assuming direct $CP$ violation to be negligible,  it can be shown~\cite{STOOF} that these observables have the following dependence on the underlying physics parameters
\begin{eqnarray}
A_\Gamma & \approx & \frac{1}{2} y\cos\phi_{CP} \left(r_{CP} - \frac{1}{r_{CP}}  \right) -  \nonumber  \\
& & \frac{1}{2}x\sin\phi_{CP} \left(\frac{1}{r_{CP}} + r_{CP}\right),  \label{eq:ag_eigen}\\
y_{CP} & \approx &  \frac{1}{2} y\cos\phi_{CP} \left( \frac{1}{r_{CP}} +  r_{CP}\right)  -  \nonumber \\
& & \frac{1}{2} x\sin\phi_{CP} \left(r_{CP} - \frac{1}{r_{CP}}   \right). \label{eq:ycp_eigen}
\end{eqnarray}
Expressions that also allow for the contribution of direct $CP$ violation can be found in Ref.~\cite{MARCO}.
Thus in the limit of $CP$ conservation $A_\Gamma$ vanishes and $y_{CP} \rightarrow y$.  The average of currently available measurements, dominated by studies based on the $CP$-even eigenstates $K^+K^-$ and $\pi^+\pi^-$, yields $A_\Gamma = (-0.058 \pm 0.040) \%$ and $y_{CP}= (0.866 \pm 0.155 ) \%$~\cite{HFAG}.  (Here the $A_\Gamma$ average includes new measurements from the LHCb~\cite{AG_LHCBSEMILEP} and CDF~\cite{AG_CDF} collaborations, in addition to the older set of results from LHCb~\cite{AG_LHCBPROMPT}, BaBar~\cite{AG_BABAR} and Belle~\cite{AG_BELLE} that are considered in Ref.~\cite{HFAG}.)

\vspace{0.2cm}
\noindent
{\bf Introducing  self-conjugate multibody decays and the  \boldmath $CP$-even fraction $F_+$} 
\vspace{0.2cm}

The $CP$ content of an inclusive self-conjugate multibody decay, for example $D \to \pi^+\pi^-\pi^0$,  can be measured with a sample of  coherently produced $D\Dbar$ pairs at the $\psi(3770)$ resonance, such as that collected by the CLEO-c and BESIII experiments.  A double-tag technique is employed in which one $D$ meson is reconstructed in the signal decay of interest, and the other in its decay to a $CP$ eigenstate.  In such an event, and neglecting any $CP$ violation, the quantum numbers of the $\psi(3770)$ meson  means that the $CP$ eigenvalue of the signal decay is fixed.   The $CP$-even fraction of the signal decay is given by $F_+ = N^+ / (N^+ + N^-)$, where $N^+$ ($N^-$) designates the number of decays tagged as $CP$-even (-odd), after correction for detector inefficiencies and the specific branching fractions of the $CP$ eigenstate tags employed.  In this manner $F_+$ has been measured for the decay $D \to \pi^+\pi^-\pi^0$ and found to be $0.973 \pm 0.017$, indicating the mode to be almost fully $CP$ even~\cite{MINAKSHI,FOURPI}.

Although $CP$ violation is neglected in the currently available measurements of $F_+$  this assumption introduces negligible bias in the result.  
Both the Standard Model and  theories of New Physics  expect direct $CP$ violation in charm decays to be  $\le 10^{-3}$~\cite{THEORY2}, a
prediction which is compatible with existing experimental results~\cite{PDG}.  Any effects will therefore be small alongside the measurement precision attainable with the CLEO-c and current BESIII data sets.  Furthermore, the double-tag analyses performed at these experiments have no sensitivity to indirect $CP$ violation at leading order in $(x,y)$, as the $D\Dbar$ system is produced at rest.  For the specific case of $D \to \pi^+\pi^-\pi^0$, a recent time-integrated high precision analysis by LHCb has revealed no evidence of any direct $CP$-violating effects~\cite{PIPIPI0}. 

There is a simple relationship between $F_+$ and the parameters that describe the intensity and strong-phase variation over the phase space of the decay.  
The amplitude of a multibody decay such as $D \to \pi^+\pi^-\pi^0$ is dependent on the final-state kinematics, which can be uniquely defined by 
the Dalitz plot coordinates $s_{12}=m^2(\pi^+\pi^0)$ and $s_{13}=m^2(\pi^-\pi^0)$. 
The amplitude of a $D^0$ decay to a specific final state is given by $\mathcal{A}_{D^0}(s_{12},s_{13}) = a_{12,13}e^{i\delta_{12,13}}$, where the integral of $|\mathcal{A}_{D^0}(s_{12},s_{13})|^2$ over the full Dalitz plot is normalised to unity.
Consider the situation where the  Dalitz plot is divided into two bins by the line $s_{12} = s_{13}$.
The bin for which $s_{12} > s_{13}$ is labelled $-1$ and the opposite bin is labelled $+1$.
The parameter $K_i$ ($\Kbar_i$) is the flavour-tagged fractional intensity, being the proportion of decays to fall in bin $i$ in the case that the mother particle  is known to be a $D^0$ (\Dzb) meson:
\begin{equation}
K_i \equiv  \int_i |a_{12,13}|^2\, \mathrm{d}s_{12}\, \mathrm{d}s_{13}.
\end{equation} 
The parameter $c_i$ is the cosine of the strong-phase difference between $D^0$ and \Dzb decays averaged in bin $i$ and weighted by the absolute decay rate:
\begin{equation}
c_i \equiv  \int_i \frac{a_{12,13}\bar{a}_{12,13} \cos(\delta_{12,13} - \bar{\delta}_{12,13})}{\sqrt{K_i \Kbar_{i}}}\, \mathrm{d}s_{12}\, \mathrm{d}s_{13}.
\end{equation}
A parameter $s_i$ is defined in an analogous manner for the sine of the strong-phase difference.  

The $CP$-tagged populations of these bins, $N_{i}^\pm$, normalised by the corresponding single $CP$-tag yields, is given by~\cite{ANTONBONDAR}
\begin{eqnarray}
N_i^\pm & = & h_D \Big(K_i \, \pm \, 2 c_i \sqrt{K_i \Kbar_{i}} \, + \, \Kbar_{i} \Big).
\label{eq:nbins}
\end{eqnarray}
Here $h_D$ is a normalisation factor independent of bin number and $CP$ tag. 
When there is no direct $CP$ violation in the decay  $\mathcal{A}_{\Dzb}(s_{12},s_{13}) = \bar{a}_{12,13}e^{i\bar{\delta}_{12,13}} \equiv a_{13,12}e^{i\delta_{13,12}}  $  and so
\begin{eqnarray}
\Kbar_i = {K_{-i}}, & c_i = c_{-i}\; {\rm and} \;  s_i = -s_{-i}.
\label{eq:cprelations}
\end{eqnarray}
Under this assumption, and the identities $N^{\pm} = \sum_{i} N_i^\pm$, and $\sum_i K_i = 1$, it follows that in the two-bin case
\begin{equation}
F_+ = \frac{1}{2} \left( 1 \, + \, 2 c_1 \sqrt{K_1 K_{-1}} \right).
\label{eq:fplusbondar}
\end{equation}

\vspace{0.2cm}
\noindent
{\bf \boldmath Measurements with  inclusive self-conjugate multibody decays} 
\vspace{0.2cm}

Now consider, for an incoherently produced $D$ meson, the time dependence of a self-conjugate multibody decay.
The time evolution of the $D^0$ to the point $(s_{12},s_{13})$ is given by
\begin{eqnarray}
\mathcal{A}_{D^0}(t,s_{12},s_{13})& = & a_{12,13}e^{i\delta_{12,13}}g_{+}(t) \nonumber \\
 & +& r_{CP}e^{i\phi_{CP}}a_{13,12}e^{i\delta_{13,12}}g_{-}(t),
\end{eqnarray} 
where $g_\pm(t)=\frac{1}{2}[e^{-i(M_1 - i\Gamma_1/2)t} \pm e^{-i(M_2 - i\Gamma_2/2)t}]$.
Ignoring terms of $\mathcal{O}(x^2,y^2,xy)$ or higher, the rate of decay to that point is proportional to
\begin{eqnarray}
|\mathcal{A}_{D^0}(t,s_{12},s_{13})|^2& =& e^{-\Gamma t} \bigg\{ a_{12,13}^2 - a_{12,13}a_{13,12}r_{CP}\Gamma t \times \nonumber \\
& & \Big[ y \cos (\delta_{12,13} - \delta_{13,12} - \phi_{CP}) + \nonumber \\
& & x \sin (\delta_{12,13} - \delta_{13,12} - \phi_{CP})\Big]\bigg \}. \nonumber \\
& & 
\end{eqnarray}
Integrating this over the two bins of the full Dalitz plot leads to the time-dependent decay probability
\begin{eqnarray}
\mathcal{P}(D^0(t)) &=& \int_{+1} | \mathcal{A}_{D^0}(t,s_{12},s_{13})|^2\, \mathrm{d}s_{12}\, \mathrm{d}s_{13} + \nonumber \\
& & \int_{-1} | \mathcal{A}_{D^0}(t,s_{12},s_{13})|^2\, \mathrm{d}s_{12}\, \mathrm{d}s_{13} \\
  &=&{\rm exp}{(-\Gamma t)}\Big[1 -  r_{CP}(2F_+-1)\times \nonumber \\
&& (y\cos\phi_{CP} - x\sin\phi_{CP}) \Gamma t \Big],
\end{eqnarray}
where use is made of the definitions of $c_i,s_i$ and the relations given in Eqs.~(\ref{eq:cprelations}) and~(\ref{eq:fplusbondar}).
Hence the width of the decay is approximated at first order in $x$ and $y$ by
\begin{equation}
\hat{\Gamma} \approx \Gamma\Big[1 +  r_{CP}(2F_+-1)\times(y\cos\phi_{CP} - x\sin\phi_{CP})\Big].
\end{equation} 
The time evolution for the \Dzb decay to the point $(s_{12},s_{13})$ is given by
\begin{eqnarray}
\mathcal{A}_{\Dzb}(t,s_{12},s_{13})& = & \frac{1}{r_{CP}}e^{-i\phi_{CP}}a_{12,13}e^{i\delta_{12,13}}g_{-}(t) \nonumber \\
 & +& a_{13,12}e^{i\delta_{13,12}}g_{+}(t),
\end{eqnarray}
and thus the width of the \Dzb decay is approximated by
\begin{equation}
\hat{\bar{\Gamma}} \approx \Gamma\Big[1 + \frac{1}{r_{CP}}(2F_+-1) \times (y\cos\phi_{CP} + x\sin\phi_{CP}) \Big] .
\end{equation} 
Defining $A_\Gamma^{\rm eff} \equiv  \frac{ \hat{\Gamma} - \hat{\bar{\Gamma}} }{\hat{\Gamma} + \hat{\bar{\Gamma}}}$ and $y_{CP}^{\rm eff} \equiv \left( \frac{\hat{\Gamma} + \hat{\bar{\Gamma}} }{2\Gamma}  - 1 \right)$ it follows that 
\begin{eqnarray}
A_\Gamma^{\rm eff}  &\approx& \frac{1}{2}  (2F_+-1)y \cos\phi_{CP} \left(r_{CP} - \frac{1}{r_{CP}}  \right) -  \nonumber  \\
&& \frac{1}{2}(2F_+-1)x \sin\phi_{CP} \left(r_{CP}+ \frac{1}{r_{CP}}\right), \label{eq:aeff}\\
y_{CP}^{\rm eff} &\approx&   \frac{1}{2} (2F_+-1)y \cos\phi_{CP} \left( r_{CP}+\frac{1}{r_{CP}} \right)  -  \nonumber \\
&& \frac{1}{2} (2F_+-1)x \sin\phi_{CP} \left(r_{CP} - \frac{1}{r_{CP}}   \right)\label{eq:yeff}.
\end{eqnarray}

These expressions contain an additional dilution factor of $(2F_+ -1)$ in comparison to the $CP$-eigenstate relations of Eqs.~(\ref{eq:ag_eigen}) and~ (\ref{eq:ycp_eigen}) and are identical in the case when $F = 0$ or $1$.  In the limit $F_+ \to 0.5$  then both observables vanish. It is interesting to note that a similar relationship between the two classes of $D$ decays was found in Ref.~\cite{MINAKSHI} when considering the determination of the unitarity triangle angle $\gamma$ using  $B^\pm \to D K^\pm$ decays.

Expressions~(\ref{eq:aeff}) and~(\ref{eq:yeff}) may be modified to allow for the possible contribution of direct $CP$ violation.
In this case the relations in Eq.~(\ref{eq:cprelations}) no longer apply. Direct $CP$ violation adds an additional magnitude and weak phase difference when considering the relations between the amplitude of the $D^0$ and \Dzb decay, and this additional magnitude and phase varies as a function of position in phase space.

With the inclusion of direct $CP$ violation the expression for $A_{\Gamma}^{\rm eff}$ becomes 
\begin{eqnarray}
A_{\Gamma}^{\rm eff} &\approx& \frac{1}{2} \Bigg[ (2F_+^{\prime}-1)y \cos\phi_{CP} \left(r_{CP} - \frac{1}{r_{CP}}  \right) -  \nonumber  \\
 & &(2F_+^{\prime}-1)x \sin\phi_{CP} \left(r_{CP} + \frac{1}{r_{CP}} \right) + \nonumber \\
&& y \Delta  \sin \phi_{CP}\left(r_{CP} - \frac{1}{r_{CP}}  \right) + \nonumber \\ 
&& x \Delta  \cos \phi_{CP}\left(r_{CP} + \frac{1}{r_{CP}}  \right) \Bigg],
\label{eq:agwithdirect}
\end{eqnarray} 
where $r_{CP}$ and $\phi_{CP}$ are unchanged in their meaning and relate only to indirect $CP$ violation, $(2F_+'-1)  \equiv c_1\sqrt{K_1\Kbar_1} + c_{-1}\sqrt{K_{-1}\Kbar_{-1}}$ and $\Delta \equiv s_1\sqrt{K_1\Kbar_1} + s_{-1}\sqrt{K_{-1}\Kbar_{-1}}$.  Hence the effect of the additional amplitudes due to direct $CP$ violation is contained within the terms $F_+^{\prime}$ and $\Delta$. 
In the limit of no direct $CP$ violation $\Delta \to 0$, and $F_+^{\prime} \to F_+$. Since $\Delta$ must be small the third term in Eq.~(\ref{eq:agwithdirect}) is negligible in comparison to the others.

The expression for $y_{CP}^{\rm eff}$ becomes
\begin{eqnarray}
y_{CP}^{\rm eff}   &\approx&   \frac{1}{2} \Bigg[ (2F_+^{\prime}-1)y\cos\phi_{CP} \left(r_{CP} + \frac{1}{r_{CP}}\right)  -  \nonumber \\
&&  (2F^{\prime}_+-1)x\sin\phi_{CP} \left(r_{CP} - \frac{1}{r_{CP}} \right) + \nonumber\\
&&  y \Delta  \sin\phi_{CP} \left(r_{CP} + \frac{1}{r_{CP}}\right) + \nonumber\\
&&   x \Delta  \cos\phi_{CP} \left(r_{CP} - \frac{1}{r_{CP}}\right) \Bigg].
\end{eqnarray}

\vspace{0.2cm}
\noindent
{\bf Discussion and conclusions}
\vspace{0.2cm}

Measurements of $A_\Gamma^{\rm eff}$ and $y_{CP}^{\rm eff}$ performed with any self-conjugate multibody decay can be used to determine $A_\Gamma$ and $y_{CP}$, respectively, provided that  the $CP$ content of the decay is known.
The mode $D \to \pi^+\pi^-\pi^0$ is a very promising candidate for this purpose  since the dilution effects arising from the factor $(2F_+ -1)$ in Eqs.~(\ref{eq:aeff}) and~(\ref{eq:yeff}) are $<10\%$, and it possesses a branching fraction that is around 3.5 times  higher  than   that of $D \to K^+K^-$, the most common $CP$-eigenstate mode used for these measurements. Therefore this channel offers an opportunity to improve the knowledge of $A_\Gamma$ and $y_{CP}$ significantly, particularly at $e^+e^-$ experiments such as Belle-II, where the $\pi^0$ reconstruction efficiency is good.   The relatively abundant four-body decay $D \to \pi^+\pi^-\pi^+\pi^-$, which has a  $CP$-even fraction of $0.737 \pm 0.028$~\cite{FOURPI}, also has the potential to be a high impact channel.  The sensitivities of these channels are compared to those of the established $CP$-eigenstate decays, $D \to K^+K^-$ and $D \to \pi^+\pi^-$,  in Table~\ref{tab:sensitivity}, assuming the same trigger and reconstruction efficiency for all.

\begin{table}[tbh]
\begin{center}
\caption{Relative uncertainties on $A_\Gamma$ and $y_{CP}$ for the multibody modes $D \to \pi^+\pi^-\pi^0$ and $D \to \pi^+\pi^-\pi^+\pi^-$ compared with those of the  $CP$ eigenstate modes $D \to K^+K^-$ and $D \to \pi^+\pi^-$, assuming the measured central values of the branching fractions ($BF$)~\cite{PDG} and $CP$-even fractions ($F_+$)~\cite{FOURPI}. The uncertainties are all normalised to that of $D \to K^+K^-$.} \vspace*{0.1cm} \label{tab:sensitivity}
\begin{tabular}{lcccc} \hline\hline 
& \hspace{0.1cm}$K^+K^-$\hspace{0.1cm} & \hspace{0.1cm}$\pi^+\pi^-$\hspace{0.1cm} & \hspace{0.1cm}$\pi^+\pi^-\pi^0$\hspace{0.1cm} & \hspace{0.1cm}$\pi^+\pi^-\pi^+\pi^-$\hspace{0.1cm} \\ \hline
$BF$ $[\times 10^{-2}]$ & $0.396$ & $0.1402$ &  $1.43$  &  $0.742$ \\
$F_+$ & 1 & 1 &  0.973  & 0.737 \\
Uncertainty & 1 & 1.68 & 0.56 & 1.54 \\ \hline\hline
\end{tabular}
\end{center}
\end{table}

Another mode of potential interest is $D \to K^0_{\rm S} \pi^+\pi^-\pi^0$, which has a branching fraction of over $5\%$ and comprises the $CP$-odd eigenstates $K^0_{\rm S} \eta$ and $K^0_{\rm S} \omega$ as submodes, although its sensitivity cannot be assessed until its $CP$ content is measured.  This channel also has the feature of being Cabibbo favoured, which means that it is extremely robust against any pollution from direct $CP$ violation.  The extensively studied decay $D \to K^0_{\rm S} \pi^+\pi^-$ is not suitable for an inclusive treatment, since it has a $CP$ content of $F_+ \sim 0.5$, as is evident from examining the relative proportion of $CP$-even and $CP$-odd double-tagged events reported in a CLEO analysis performed to measure the $c_i$ and $s_i$ parameters~\cite{CLEOKSPIPI}.

The Belle collaboration has reported a model-dependent analysis of the mode $D \to K^0_{\rm S}K^+K^-$ that measures $y_{CP}$ through comparing the $CP$-odd and $CP$-even regions of the Dalitz plot~\cite{BELLEKSKK}.  Studies also exist that fit time-dependent amplitude models to the Dalitz plots of the decays $D \to K^0_{\rm S} \pi^+ \pi^-$ and
$D \to K^0_{\rm S} K^+ K^-$ in order to determine the mixing and $CP$-violation parameters~\cite{KSHHEXP1,KSHHEXP2,KSHHEXP3}.  Furthermore, proposals have been made of how to perform model-independent analyses of self-conjugate decays binned in phase space~\cite{KSHHTH1,KSHHTH2}.   The method advocated in this paper is novel because it is inclusive, model independent and suitable for those decays which are dominated by a single $CP$ eigenstate, such as $D \to \pi^+\pi^-\pi^0$.  Inclusive analyses are experimentally more straightforward since there is no need to account for the position in phase space of each decay, provided that the acceptance is relatively uniform.

As explained in Ref.~\cite{MINAKSHI}, self-conjugate multibody modes can also be used to measure the unitarity triangle angle $\gamma$ with $B^\pm \to DK^\pm$ decays as long as $F_+$ is known for the mode under consideration.  In cases where no measurement of $F_+$ exists from the charm threshold it is possible to obtain this information from a comparison of a measurement of $y_{CP}^{\rm eff}$ and the value of $y_{CP}$ obtained from $CP$ eigenstates, or indeed that of $y$ itself, assuming negligible $CP$ violation in the charm system. This strategy of using charm-mixing observables to help provide input for the $\gamma$ determination is similar to that already proposed for quasi-flavour specific states~\cite{SAMJONAS}.

In summary, inclusive measurements of the time evolution of mutibody self-conjugate charm decays offer the possibility to obtain significantly improved sensitivity to  $CP$ violation and mixing in the $D^0\Dzb$ system. The observables $A_\Gamma^{\rm eff}$ and $y^{\rm eff}_{CP}$ are simply related to those of the $CP$ eigenstate case,  $A_\Gamma$ and $y_{CP}$, by a dilution factor $(2F_+ - 1)$,  where $F_+$ is the fractional $CP$-even content of the decay.  This parameter may be measured in coherently  produced $D\Dbar$   decays at the $\psi(3770)$.   One of the modes for which $F_+$  is known, $D \to \pi^+\pi^-\pi^0$, has the potential to yield a more precise determination of $A_\Gamma$ and $y_{CP}$ than is possible with $CP$ eigenstate decays, and another, $D \to \pi^+\pi^-\pi^+\pi^-$, also offers good sensitivity.  Other promising channels exist with relatively high branching fractions and should also be exploited, provided that analyses at the $\psi(3770)$ show them  to be dominated by a single $CP$ eigenstate.   Alternatively, measurements of $y_{CP}^{\rm eff}$ using these latter channels will allow their $CP$ content to be determined, which is valuable input for the programme to measure the unitarity angle $\gamma$.  First results using this class of decays are eagerly awaited.

\vspace{0.2cm}
\noindent
{\bf Acknowledgments}
\vspace{0.2cm}

We thank Jim Libby for useful discussions out of which the idea for this paper emerged.  We are grateful to Tim Gershon and Marco Gersabeck for valuable suggestions.
We acknowledge support from the UK Science and Technology Facilities Council and the UK-India Education and Research Initiative.

%



\begin{thebibliography}{99}

\bibitem{THEORY1} M.~Bobrowski, A.~Lenz, J.~Riedl and J.~Rohrwild, {\it How large can the SM contribution to $CP$ violation in $D^0-\Dzb$ mixing be?}, JHEP {\bf 03} (2010) 009, arXiv:1002.4794.
\bibitem{THEORY1b} M.~Ciuchini {\it et al.}, {\it $D-\Dbar$ mixing and new physics: general considerations and constraints on the MSSM}, Phys Lett. {\bf B 655} (2007) 162, arXiv:hep-ph/0703204.
\bibitem{MINAKSHI} M.~Nayak {\it et al.}, {\it First determination of the $CP$ content of $D \to \pi^+\pi^-\pi^0$ and $D \to K^+K^-\pi^0$}, Phys. Lett. {\bf B 740} (2015) 1, arXiv:1410.3964 [hep-ex].
\bibitem{FOURPI}  S.~Malde {\it et al.}, {\it First determination of the $CP$ content of $D \to\pi^+\pi^-\pi^+\pi^-$ and updated determination of the $CP$ contents of 
$D\to\pi^+\pi^-\pi^0$ and  $D\to K^+K^-\pi^0$}, arXiv:1504.05878 [hep-ex].
\bibitem{HFAG} Y.~Amhis {\it et al.} (HFAG), {\it Averages of b-hadron, c-hadron, and $\tau$-lepton properties as of summer 2014}, arXiv:1412.7515 [hep-ex], online updates at http://www.slac.stanford.edu/xorg/hfag. 
\bibitem{STOOF} S.~Bergmann, Y.~Grossman, Z.~Ligeti, Y.~Nir and A.A.~Petrov, {\it Lessons from CLEO and FOCUS measurements of $D^0-\Dzb$ mixing parameters}, Phys. Lett. {\bf B 486} (2000) 418, arXiv:hep-ph/0005181.
\bibitem{MARCO} M.~Gersabeck {\it et al.}, {\it On the interplay of direct and indirect $CP$ violation in the charm sector}, J. Phys. {\bf G 39} (2012) 045005, arXiv:1111.6515 [hep-ex].
\bibitem{AG_LHCBSEMILEP} R.~Aaij {\it et al.} (LHCb collaboration), {\it Measurement of indirect $CP$ asymmetries in $D^0 \to K^-K^+$ and $D^0 \to \pi^-\pi^+$ decays}, JHEP {\bf 04} (2015) 043, arXiv:1501.06777 [hep-ex].
\bibitem{AG_CDF} T.A.~Aaltonen {\it et al.} (CDF collaboration), {\it Measurements of indirect $CP$-violating asymmetries in $D^0 \to K^+K^-$ and $D^0 \to \pi^+\pi^-$ decays at CDF}, Phys. Rev. {\bf D 90} (2014) 111103, arXiv:1410.5435 [hep-ex].
\bibitem{AG_LHCBPROMPT}R.~Aaij {\it et al.} (LHCb collaboration), {\it Measurements of indirect $CP$ asymmetries in $D^0 \to K^-K^+$ and $D^0 \to \pi^+\pi^-$ decays}, Phys. Rev. Lett. {\bf 112} (2014) 041801, arXiv:1310.7201 [hep-ex]. 
\bibitem{AG_BABAR} J.P.~Lees {\it et al.} (BaBar collaboration), {\it Measurement of $D^0-\Dzb$ mixing and $CP$ violation in two-body $D^0$ decays}, Phys. Rev. {\bf D 87} (2013) 012004, arXiv:1209.3896 [hep-ex].
\bibitem{AG_BELLE} M.~Staric (for the Belle collaboration), {\it New Belle results on $D^0-\Dzb$ mixing}, arXiv:1212.3478 [hep-ex].
\bibitem{THEORY2} Y.~Grossman, A.L.~Kagan and Y.~Nir, {\it New physics and $CP$ violation in singly Cabibbo suppressed $D$ decays}, Phys. Rev. {\bf D 75} (2007) 036008, arXiv:hep-ph/0609178.
\bibitem{PDG} K.A.~Olive {\it et al.} (Particle Data Group), {\it Review of particle physics}, Chin. Phys. {\bf C 38} (2014) 090001.
\bibitem{PIPIPI0} R.~Aaij {\it et al.} (LHCb collaboration), {\it  Search for $CP$ violation in $D^0 \to \pi^+\pi^-\pi^0$ decays with the energy test}, Phys. Lett. {\bf B 740} (2015) 158, arXiv:1410.4170 [hep-ex].
\bibitem{ANTONBONDAR} A.~Bondar and A.~Poluektov, {\it The use of quantum correlated $D^0$ decays for $\phi_3$ measurement}, Eur. Phys. J. {\bf C 55} (2008) 51, arXiv:0801.0840 [hep-ex].
\bibitem{CLEOKSPIPI} J.~Libby {\it et al.} (CLEO collaboration), {\it Model-independent determination of the strong-phase difference between the decays $D^0$ and $\Dzb \to K^0_{S,L} h^+h^- $ ($h=\pi, K)$ and its impact on the measurement of the CKM angle $\gamma$}, Phys. Rev. {\bf D 82} (2010) 112006, arXiv:1010.2817 [hep-ex].
\bibitem{BELLEKSKK} A.~Zupanc {\it et al.} (Belle collaboration), {\it Measurement of $y_{CP}$ in $D^0$ meson decays to the $K^0_S K^+K^-$ final state}, Phys. Rev. D {\bf 80} 052006 (2009), arXiv:0905.4185 [hep-ex].
\bibitem{KSHHEXP1} P. del Amo Sanchez {\it et al.} (BaBar collaboration), {\it Measurement of the $D^0-\Dzb$ mixing parameters using $D^0 \to K_S^0 \pi^+\pi^-$ and $D^0 \to K_S^0 K^+ K^-$ decays}, Phys. Rev. Lett. {\bf 105} (2010) 081803, arXiv:1004.5053 [hep-ex].
\bibitem{KSHHEXP2}
T.~Peng {\it et al.} (Belle collaboration), {\it Meausurement of  $D^0-\Dzb$ mixing and search for indirect CP violation using $D^0 \to K_S^0 \pi^+\pi^-$ decays}, Phys. Rev.  {\bf D 89} (2014)  091103(R), arXiv:1404.2412 [hep-ex].
\bibitem{KSHHEXP3} D.M.~Asner {\it et al.} (CLEO collaboration), {\it Search for $D^0 \Dzb$ mixing in the Dalitz plot analysis of $D^0 \to K_S^0\pi^+\pi^-$}, Phys. Rev. {\bf D 72} (2005) 012001, arXiv:hep-ex/0503045.
\bibitem{KSHHTH1} A.~Bondar, A.~Poluektov and V.~Vorobiev, {\it Charm mixing in the model-independent analysis of correlated $D^0\Dzb$ decays}, Phys. Rev. {\bf D 82} (2010) 034033, arXiv:1004.2350 [hep-ph].
\bibitem{KSHHTH2} C.~Thomas and G.~Wilkinson, {\it Model-independent $D^0-\Dzb$ mixing and CP violation studies with $D^0 \to K_S^0 \pi^+\pi^-$ and $D^0 \to K_S^0 K^+ K^-$}, JHEP {\bf 10} (2012) 185, arXiv:1209.0172 [hep-ex].
\bibitem{SAMJONAS} S.~Harnew and J.~Rademacker, {\it Charm mixing as input for model-independent determinations of the CKM phase $\gamma$},  Phys. Lett. {\bf B 728} (2014) 296, arXiv:1309.0134 [hep-ph].


\end{thebibliography}
\end{document}